\documentclass[showpacs,preprint,amsmath,amssymb,nofootinbib]{revtex4}

\usepackage{color}
\usepackage{graphicx}
\usepackage{dcolumn}
\usepackage{bm}
\usepackage{lscape}
\usepackage{ulem}

\begin{document}
\title{Medium effect in high-density nuclear matter probed by systematic analyses of nucleus-nucleus elastic scattering}

\author{T.~Furumoto}
\email{furumoto-takenori-py@ynu.ac.jp}
\affiliation{Graduate School of Education, Yokohama National University, Yokohama 240-8501, Japan}
\affiliation{National Institute of Technology, Ichinoseki College, Ichinoseki, Iwate 021-8511, Japan}

\author{Y.~Sakuragi}%
\affiliation{Department of Physics, Osaka City University, Osaka 558-8585, Japan}

\author{Y.~Yamamoto}
\affiliation{RIKEN Nishina Center, RIKEN, Wako, Saitama 351-0198, Japan}

\date{\today}

\begin{abstract}
We investigate the property of the high-density nuclear matter probed by the nucleus-nucleus elastic scattering in the framework of the double-folding (DF) model with the complex $G$-matrix interaction.
The medium effect including three-body-force (TBF) effect is investigated with present two methods based on the frozen density approximation (FDA).
The medium effect is clearly seen on the potential and the elastic cross section for the $^{16}$O~+~$^{16}$O system at $E/A =$ 70 MeV.
The crucial role of the medium effect is also confirmed with other effective nucleon-nucleon ($NN$) interactions.
In addition, the present methods are applied to other heavy-ion elastic scattering systems.
Again, the medium effect is clearly seen in the heavy-ion elastic cross section.
The medium effect on the elastic cross section becomes invisible with the increase of the target mass and the incident energy (up to $E/A =$ 200 MeV).
However, the medium effect is again important to fix the heavy-ion scattering over $E/A =$ 200 MeV.
Finally, we make clear the crucial role of the TBF effect up to $k_F =$ 1.6 fm$^{-1}$ in the nucleus-nucleus elastic scattering.
\end{abstract}

\pacs{21.30.-x, 24.10.-i, 24.10.Ht, 25.70.Bc} 
\keywords{complex G-matrix, double-folding model, frozen-density approximation}

\maketitle

\section{Introduction}

The study of the medium effect through the nuclear structure and reaction gives the understanding of the property of the neutron star.
Many nuclear structure and reaction models are based on the nucleon-nucleon ($NN$) interaction to study the medium effect.
It is useful to investigate the colliding two nuclei on the earth in order to obtain the information of the medium effect in the high density region exceeding the normal density.
The elastic scattering has been investigated with the double-folding (DF) model based on the frozen density approximation (FDA) to understand the nuclear saturation property~\cite{KOB82,KHO97}.

The DF model has been successful to describe the nucleus-nucleus reactions from the microscopic view point~\cite{SAT79}.
The M3Y interaction which has no density dependence is used in the DF model at the first stage.
The M3Y folding model well reproduces a lot of reaction data in the most surface regions.
The effective $NN$ interaction is developed so as to include the medium effect expressed by the density dependence~\cite{KOB82,KHO97}.
The density dependence is first introduced by the phenomenological way to reproduce the nucleus-nucleus elastic scattering.
After that, the density-dependence is improved by reproducing not only the nucleus-nucleus elastic scattering but also nuclear saturation property~\cite{KHO97}.
However, such density dependent and independent effective $NN$ interactions, which are based on the M3Y interaction, have only the real part.
Therefore, the imaginary part of the reaction potential should be constructed by the phenomenological way in the nuclear reaction analysis.

In order to solve the problem, the complex $G$-matrix interaction is applied to the DF model~\cite{FUR09}.
The complex nucleus-nucleus potential is obtained not only by folding the real part of the complex $G$-matrix interaction, but also by folding the imaginary part of the complex $G$-matrix interaction.
Therefore, the ambiguity of the imaginary part is reduced.
In addition, the important role of the three-body-force (TBF) effect is made clear to reproduce the heavy-ion elastic scattering.
The TBF effect, especially for the repulsive effect, is introduced to stiff the neutron-star EOS.
The repulsive effect is taken into account by changing the vector meson mass in Refs.~\cite{ESC04-1,ESC04-2}, and upgraded by introducing the multi-Pomeron exchange potential (MPP) in Refs.~\cite{YAM13R,YAM14}.
The MPP model includes triple and quartic pomeron exchanges, and can lead to the neutron-star EOS stiff enough to reproduce a maximum star mass over $2M_{\rm soler}$. 
Recently, such a conclusion has been obtained even in Hyperon-mixed neutron-star matter \cite{YAM14,YAM16}.
Here, we note that the TBF contribution is investigated with not only our model, but also several models in the nuclear reaction and nuclear matter \cite{RAF13, HOL13,TOY15}.

In this paper, we propose a new approach to investigate the property of the high-density nuclear matter in detail.
In Refs.~\cite{KHO97,KHO07}, the incompressibility $K$ and the heavy-ion elastic scattering are concerned to probe the nuclear saturation property.
However, our motivation is to make sure the TBF effect in the high-density EOS more specifically by analyzing the heavy-ion elastic scattering.
Especially, it is important to investigate the nuclear matter property over the normal density through the experimental data on Earth.
The present methods to investigate the medium effect over the normal density are first introduced in Ref.~\cite{FUR14R}.
But, the application is shown only for the $^{16}$O + $^{16}$O elastic scattering at $E/A = 70$ MeV system.
Therefore, we apply the present methods to several systems to confirm the reliability of the present methods to probe the medium effect over the normal density in this paper.
In addition, we test several density-dependent effective $NN$ interactions whose density-dependence is derived from different situation.
Consequently, we confirm that the present methods are a powerful tool to probe the medium effect in the high-density region.
Lastly, we apply the present methods to the high-energy heavy-ion elastic scattering over $E/A =$ 200 MeV.

\section{Formalism}
\label{sec:Formalism}
The nucleus-nucleus potential is constructed from the microscopic view point through the DF model with the complex $G$-matrix interaction.
The microscopic nucleus-nucleus potential can be written as a Hartree-Fock type potential; 
\begin{eqnarray}
U&=&\sum_{i\in A_{1},j\in A_{2}}{[<ij|v_{D}|ij>+<ij|v_{\rm{EX}}|ji>]}\\
&=&U_{D}+U_{\rm{EX}},
\end{eqnarray}
where $v_{D}$ and $v_{\rm{EX}}$ are the direct and exchange parts of complex $G$-matrix interaction. 
The exchange part is a nonlocal potential in general. 
However, by the plane-wave representation for the $NN$ relative motion \cite{SIN75, SIN79}, the exchange part can be approximately localized. 
The direct and exchange parts of the localized potential are then written in the standard form of the DF potential as 
\begin{equation}
U_{D}(R)=\int{\rho_{1}(\bm{r}_1) \rho_{2}(\bm{r}_2) v_{D}(\bm{s}; \rho, E/A)d\bm{r}_1 d\bm{r}_2}, \label{eq:dfdirect}
\end{equation}
where $\bm{s}=\bm{r}_2-\bm{r}_1+\bm{R}$, and
\begin{eqnarray}
U_{\rm{EX}}(R)&=&\int{\rho_{1}(\bm{r}_1, \bm{r}_1+\bm{s}) \rho_{2}(\bm{r}_2, \bm{r}_2-\bm{s}) v_{\rm{EX}}(\bm{s}; \rho, E/A)} \nonumber \\
&&\times \exp{ \left[ \frac{i\bm{k}(R)\cdot \bm{s}}{M} \right] } d\bm{r}_1 d\bm{r}_2,
\label{eq:dfexchange}
\end{eqnarray}
where, $\bm{k}(R)$ is the local momentum for nucleus-nucleus relative motion defined by 
\begin{equation}
k^2(R)=\frac{2mM}{\hbar^2} \left\{ E_{\rm{c.m.}}-{\rm{Re}}\ U(R)-V_{\rm{Coul.}}(R) \right\}, \label{eq:kkk}
\end{equation}
where $M=A_1A_2/(A_1+A_2)$, $E_{\rm{c.m.}}$ is the center-of-mass energy, 
$E/A$ is the incident energy per nucleon, $m$ is the nucleon mass and $V_{\rm{Coul.}}$ is the Coulomb part of the potential. 
$A_{1}$ and $A_{2}$ are the mass numbers of the projectile and target, respectively. 
The exchange part is calculated self-consistently on the basis of the local energy approximation through Eq.~(\ref{eq:kkk}). 
Here, the Coulomb potential $V_{\rm{Coul.}}$ is also obtained by folding the $NN$ Coulomb potential with the proton density distributions of the projectile and target nuclei. 
The density matrix $\rho(\bm{r}, \bm{r}')$ is approximated in the same manner as in \cite{NEG72}; 
\begin{equation}
\rho (\bm{r}, \bm{r}')
=\frac{3}{k^{\rm{eff}}_{F}\cdot s}j_{1}(k^{\rm{eff}}_{F}\cdot s)\rho \Big(\frac{\bm{r}+\bm{r}'}{2}\Big), 
\label{eq:exchden}
\end{equation}
where $k^{\rm{eff}}_{F}$ is the effective Fermi momentum \cite{CAM78} defined by
\begin{equation}
k^{\rm{eff}}_{F} 
=\left\{ \left(\frac{3\pi^2}{2} \rho \right)^{2/3}+\frac{5C_{s}(\nabla\rho)^2}{3\rho^2}
+\frac{5\nabla ^2\rho}{36\rho} \right\}^{1/2}, \;\; 
\label{eq:kf}
\end{equation}
where we adopt $C_{s} = 1/4$ following Ref.~\cite{KHO01}. 
The exponential function in Eq.~(\ref{eq:dfexchange}) is approximated by the spherical Bessel function of rank 0, $j_{0} (\frac{k(R)s}{M})$, following the standard prescription~\cite{BRI77,ROO77,BRI78,CEG83,CEG07,MIN10}.

In the present calculations, we employ the FDA for the local density.
In the FDA, the density-dependent effective $NN$ interaction is assumed to feel the local density defined as the sum of densities of colliding nuclei evaluated;
\begin{equation}
\rho = \rho_{1} (\bm{r}_1) + \rho_{2} (\bm{r}_2). \label{eq:fda}
\end{equation}
The FDA has been widely used also in the standard DF model calculations~\cite{KOB82, SAT79, KHO94, KHO97, KHO01, KAT02, FUR09}.
In Ref.~\cite{FUR09}, it is confirmed that FDA is the best prescription in the case with complex $G$-matrix interaction to construct the proper nucleus-nucleus interaction that reproduces the observed scattering data. 

Here, we introduce two types of the present methods to investigate the medium effect over the normal density.
First, the local density based on the FDA which reflects the medium effect over the normal density to the DF potential.
Then, we restrict the medium effect by the following artificial cut of the evaluated local density; 
\begin{eqnarray}
\rho&=&
\left\{
  \begin{array}{cccc}
   \rho_{1}+\rho_{2} & \ldots & (\text{if \ \ }\rho_{1}+\rho_{2} < \rho_{\rm cut} ) & \\
   \rho_{\rm cut} & \ldots & (\text{if \ \ }\rho_{1}+\rho_{2} > \rho_{\rm cut} ) & ,
  \end{array}
\right. \label{eq:cut}
\end{eqnarray}
where the $\rho_{\rm cut}$ value is varied as a parameter.
We calculate the DF potentials with several $k_{\rm cut}$ values where the relation between $k_{\rm cut}$ and $\rho_{\rm cut}$ is defined by
\begin{equation}
\rho_{\rm cut} = \frac{2}{3\pi^2}k_{\rm cut}^3. \label{eq:krho}
\end{equation}
By changing the $k_{\rm cut}$ value, the medium effect in the high-density region is controlled and investigated in the potential and observable cross section.
In this paper, we call this method as ``cutting method''.

Next, we test the sensitivity of the TBF effect in the high-density region by the following prescription for the complex $G$-matrix interaction; 
\begin{eqnarray}
v(\bm{s}; \rho, E/A)&=&
\left\{
  \begin{array}{rl}
v & {\rm (with\ TBF)} \\
\ldots & (\text{if \ \ } \rho = \rho_{1}+\rho_{2} < \rho_{\rm rep.} ) \\
v & {\rm (w/o\ TBF)} \\
\ldots & (\text{if \ \ } \rho = \rho_{1}+\rho_{2} > \rho_{\rm rep.} ) \ ,
  \end{array}
\right. \label{eq:rep}
\end{eqnarray}
where the $v$ (with TBF) and $v$ (w/o TBF) are the complex $G$-matrix interactions with and without the TBF effect, respectively.
In this paper, we apply several effective $NN$ interactions to $v$ (with TBF) and  $v$ (w/o TBF).
The detail of those interactions will be introduced later.
We change the value of $\rho_{\rm rep.}$ to investigate the TBF effect in the high-density region.
We calculate the DF potentials with several $k_{\rm rep.}$ values where the relation between $k_{\rm rep.}$ and $\rho_{\rm rep.}$ is defined by
\begin{equation}
\rho_{\rm rep.} = \frac{2}{3\pi^2}k_{\rm rep.}^3. \label{eq:krho2}
\end{equation}
Namely, we replace the complex $G$-matrix interaction with the TBF effect by that without the TBF effect when the local density $\rho$ exceeds the $\rho_{\rm rep.}$ value.
By changing the $k_{\rm rep.}$ value, the TBF effect in the high-density region is investigated in the potential and observable cross section.
In this paper, we call this method as ``replacement method''.

\section{Results}
\label{sec:Results}
In the present paper, we adopt the complex $G$-matrix interactions, which is constructed from the two-body ESC08 $NN$ interaction supplemented by the TBF effect of the MPP model~\cite{ESC08-1,ESC08-2,YAM13R}.
There have been proposed three versions of the MPP model (MPa/b/c)~\cite{YAM14}, which reproduce the $^{16}$O~+~$^{16}$O angular distribution equally well but give rise to different stiffness of EOS.
We first use the MPa version giving the stiffest EOS.
The results with the other interactions will be shown later.
With the replacement method, we apply the ESC interaction which is constructed only from the two-body ESC08 $NN$ interaction.

We first investigate the medium effects in the high-density region above the saturation density, in the case of $^{16}$O~+~$^{16}$O scattering system in detail.
Because the $^{16}$O nucleus is one of the ideal nucleus for the elastic scattering.
The $^{16}$O nucleus is the double magic number which means to have no strong channel coupling effect from the excited state on the elastic cross section.
Indeed, we have confirmed the minor role of the channel coupling effect by the $^{16}$O nucleus on the $^{16}$O~+~$^{16}$O elastic scattering at $E/A = 70$ MeV~\cite{TAK10} while with the CEG07b interaction whose property will be introduced later.
In addition, the minor role of the channel coupling effect is found to be simulated by a slight change of the renormalization factor which is an adjustable parameter.
And, the detail of the factor will be introduced later.
We then apply the equivalent analysis based on the present methods to other scattering systems including $^{12}$C, $^{28}$Si, $^{40}$Ca, $^{90}$Zr, and $^{208}$Pb nuclei.

We adopt the nucleon density of the $^{16}$O nucleus calculated from the internal wave functions generated by the orthogonal condition model (OCM) by Okabe~\cite{OKABE} based on the microscopic $\alpha$~+~$^{12}$C cluster picture. 
For the $^{12}$C, $^{28}$Si, and $^{40}$Ca nuclei, we use the nucleon densities deduced from the charge densities~\cite{CDENS} extracted from electron-scattering experiments by unfolding the charge form factor of a proton in the standard way~\cite{nuclearSize}.
For the $^{90}$Zr and $^{208}$Pb nuclei, we adopt the density-dependent Hartree-Fock (DDHF) calculation by Negele~\cite{NEG70}.

\subsection{Medium effect in high-density region}
\begin{figure}[h]
\begin{center}
\includegraphics[width=12cm]{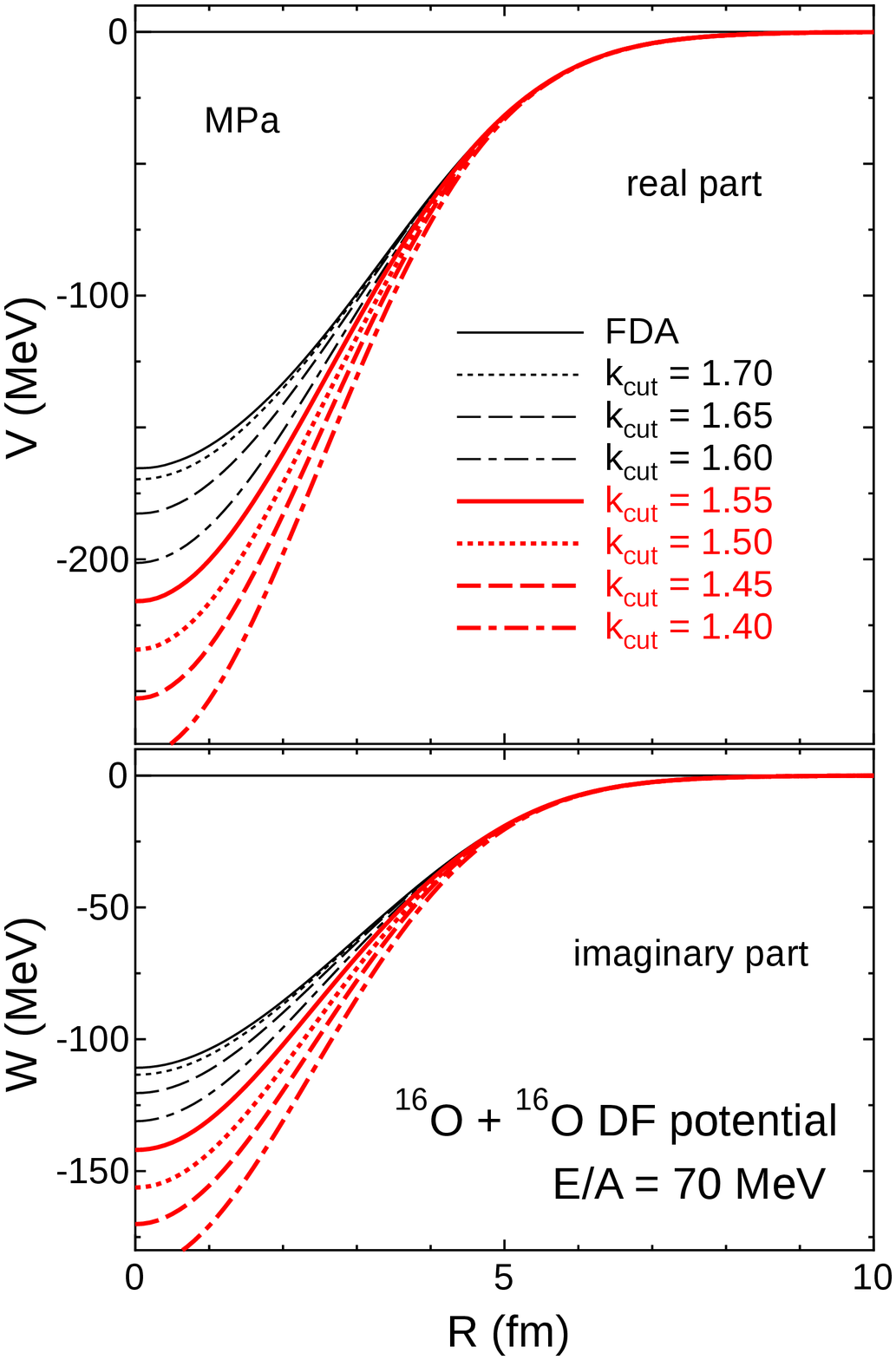}
\caption{\label{fig:01} (Color online) The real and imaginary parts of the DF potential by the cutting method (with the $k_{\rm cut}$ value). 
The solid curve is obtained by the FDA calculation.
The dotted, dashed, dot-dashed, bold-solid, bold-dotted, bold-dashed, and bold-dot-dashed curves are the results with $k_{\rm cut} = 1.70, 1.65, 1.60, 1.55, 1.50, 1.45,$ and $1.40$, respectively.}
\end{center}
\end{figure}
First, we investigate the medium effect for the high-density region with the cutting method.
In the previous work~\cite{FUR14R}, we selected the parameter as $k_{\rm cut} =$ 1.1--1.8 fm$^{-1}$ in units of 0.1 fm$^{-1}$.
In this paper, we set the parameter to be $k_{\rm cut} =$ 1.4--1.8 fm$^{-1}$ in units of 0.05 fm$^{-1}$ because one of the purpose of this work is to see medium effect in the high-density region over the normal density in detail.
Figure~\ref{fig:01} shows the real and imaginary parts of the calculated DF potential for the $^{16}$O~+~$^{16}$O elastic scattering at $E/A$ = 70 MeV.
The medium effect is clearly seen in the complex potential, especially in the inner part of the potential.
For the inner part, the local density based on the FDA reaches up to twice the normal density by approaching each other nucleus.
On the other hand, the complex $G$-matrix interaction almost feels small density in the tail region, and then, the effect around the tail part of the potential in the range of $k_{\rm cut} =$ 1.4--1.7 fm$^{-1}$ gives hardly any difference.
When the local density is restricted by Eq.~(\ref{eq:cut}), the complex $G$-matrix interaction feels the soft medium effect through the density dependence and gives the stronger potential.
In the potential, the medium effect is clearly seen up to the Fermi momentum $k_F =$ 1.70 fm$^{-1}$ (dotted curve).
Namely, it is indicated that the medium effect in the high-density region up to $k_F =$ 1.70 fm$^{-1}$ has an important role to construct the DF potential for the $^{16}$O~+~$^{16}$O system.
By way of caution, we here mention that the local density can not reach over the $k_F =$ 1.75 fm$^{-1}$ (twice the normal density) in principle because the local density is composed of the sum of the density of the colliding two nuclei as Eq.~(\ref{eq:fda}).
Then, the $k_{\rm cut}$ values are obtained up to 1.70 fm$^{-1}$.

\begin{figure}[h]
\begin{center}
\includegraphics[width=12cm]{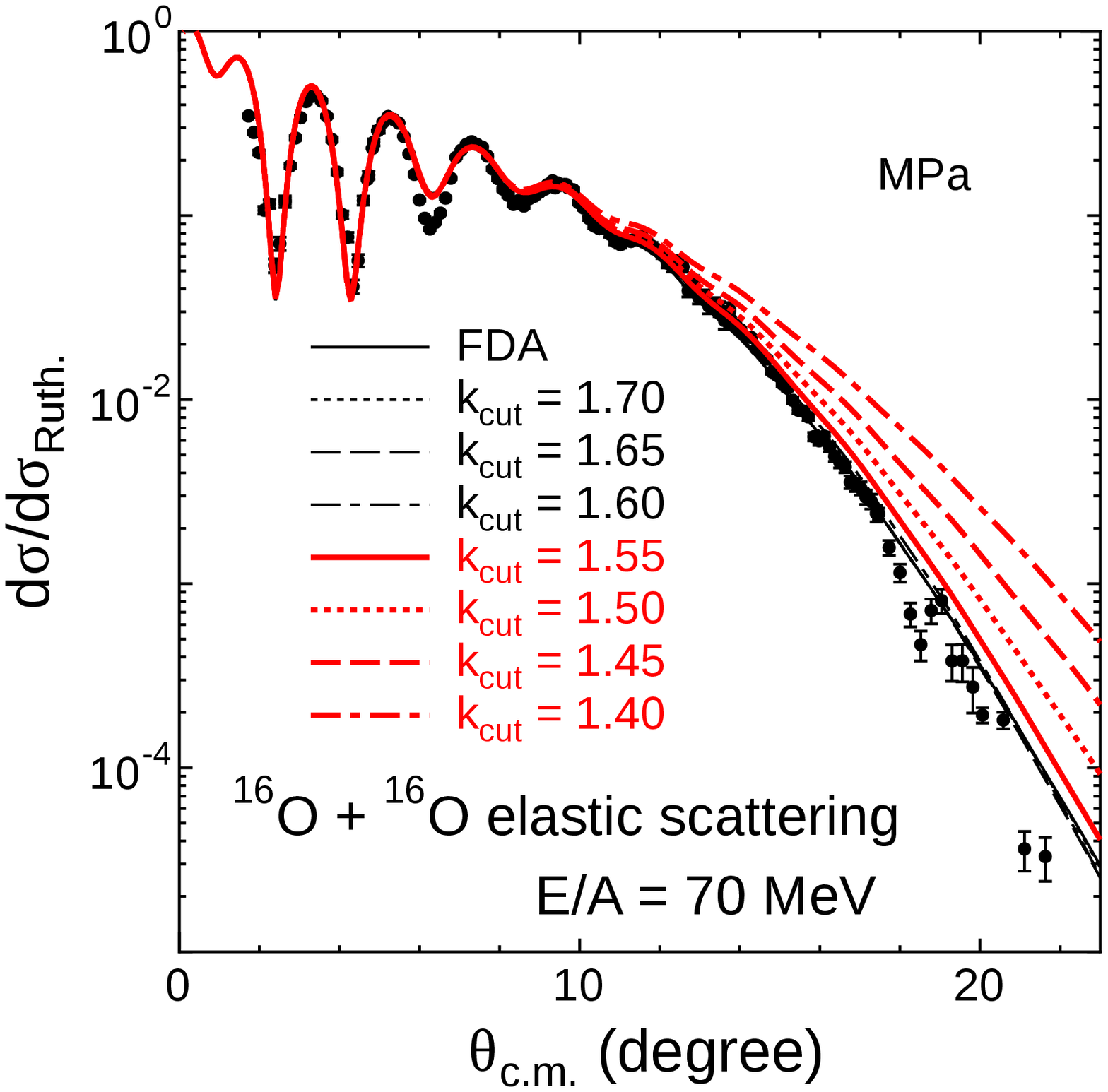}
\caption{\label{fig:02} (Color online) The elastic cross section with the DF potentials shown in Fig.~\ref{fig:01}. 
The meaning of the curves is the same as in Fig.~\ref{fig:01}.
The experimental data is taken from Ref.~\cite{NUO98}}
\end{center} 
\end{figure} 
Figure~\ref{fig:02} shows the results calculated with the DF potentials shown in Fig.~\ref{fig:01}.
The medium effect is clearly seen up to $k_F =$ 1.60 fm$^{-1}$ in the elastic cross section while the effect is seen up to $k_F =$ 1.70 fm$^{-1}$ in the potential.
The difference of the $k_F$ values is caused by the insensitivity of the most inner part of the potential for the observed elastic scattering.
These results clearly show the importance of the proper evaluation of the medium effect in the high-density region ($k_F >1.40$ fm$^{-1}$ ).
In other words, the present result implies that the medium effect in the high-density region can be probed rather sensitively through the nucleus-nucleus elastic scattering experiments at backward angles.

\begin{table}[h]
\caption{The calculated total reaction cross section for the $^{16}$O~+~$^{16}$O system at $E/A =$ 70 MeV.}
\label{tab:01}
  \begin{tabular}{cc} \hline
$k_{\rm cut}$ (fm$^{-1}$) & Total reaction cross section (mb) \\ \hline
FDA & 1428 \\ 
1.70 & 1428 \\ 
1.65 & 1428 \\
1.60 & 1429 \\ 
1.55 & 1430 \\
1.50 & 1431 \\
1.45 & 1432 \\
1.40 & 1434 \\  \hline  
  \end{tabular}
\end{table}
In addition, we calculate the total reaction cross section with the DF potentials shown in Fig.~\ref{fig:01}.
The calculated results are shown in Table~\ref{tab:01}.
The total reaction cross section almost has no sensitivity of the medium effect in the high-density region.
Namely, it is difficult to discuss the detail of the medium effect over the normal density on the total reaction cross section.

\subsection{Important role of three-body forces in high-density region}
\begin{figure}[t]
\begin{center}
\includegraphics[width=12cm]{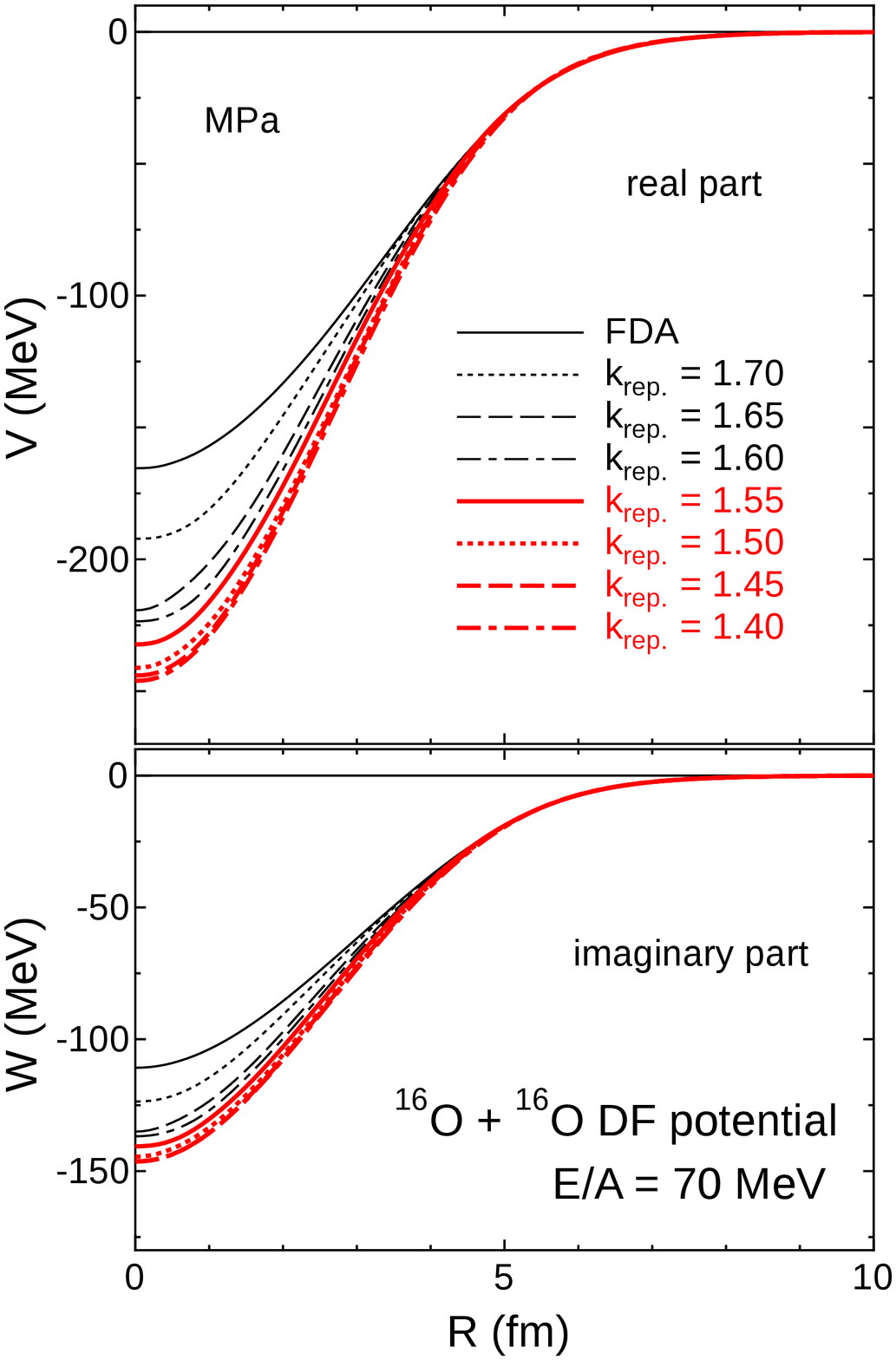}
\caption{\label{fig:03} (Color online) The real and imaginary parts of the DF potential by the replacement method (with the $k_{\rm rep.}$ value). 
The solid curve is obtained by the FDA calculation.
The dotted, dashed, dot-dashed, bold-solid, bold-dotted, bold-dashed, and bold-dot-dashed curves are the results with $k_{\rm rep.} = 1.70, 1.65, 1.60, 1.55, 1.50, 1.45,$ and $1.40$, respectively.}
\end{center} 
\end{figure}
In the previous section, the medium effect in the high-density region is clearly seen in the potential and elastic cross section.
However, the medium effect uncovered by cutting method includes both of the contributions obtained by the $G$-matrix calculation and the TBF effect.
Then, we focus on the role of the TBF effect in the high-density region with the replacement method.
We again mention that we selected the parameter as $k_{\rm rep.} =$ 1.1--1.8 fm$^{-1}$ in units of 0.1 fm$^{-1}$ in the previous work~\cite{FUR14R}.
In this paper, we set the parameter to be $k_{\rm rep.} =$ 1.4--1.8 fm$^{-1}$ in units of 0.05 fm$^{-1}$.
Figure~\ref{fig:03} shows the real and imaginary parts of the calculated DF potential for the $^{16}$O~+~$^{16}$O elastic scattering at $E/A$ = 70 MeV.
The TBF effect is clearly seen for the complex potential, especially for the inner part of the potential.
In the potential, the TBF effect is clearly seen over $k_F =$ 1.70 fm$^{-1}$.
Especially, the TBF effect over $k_F =$ 1.65 fm$^{-1}$ is remarkable for the inner part of the potential, because the effect of the three-body repulsive force is known to be important in the high-density region.
This result indicates the importance of the TBF effect in the high-density region.
When the $k_{\rm rep.}$ value becomes small, the obtained DF potential closes to the DF potential with the ESC interaction (w/o the TBF effect).
This implies that the TBF effect up to normal density is petty for the inner part of the DF potential.

\begin{figure}[t]
\begin{center}
\includegraphics[width=12cm]{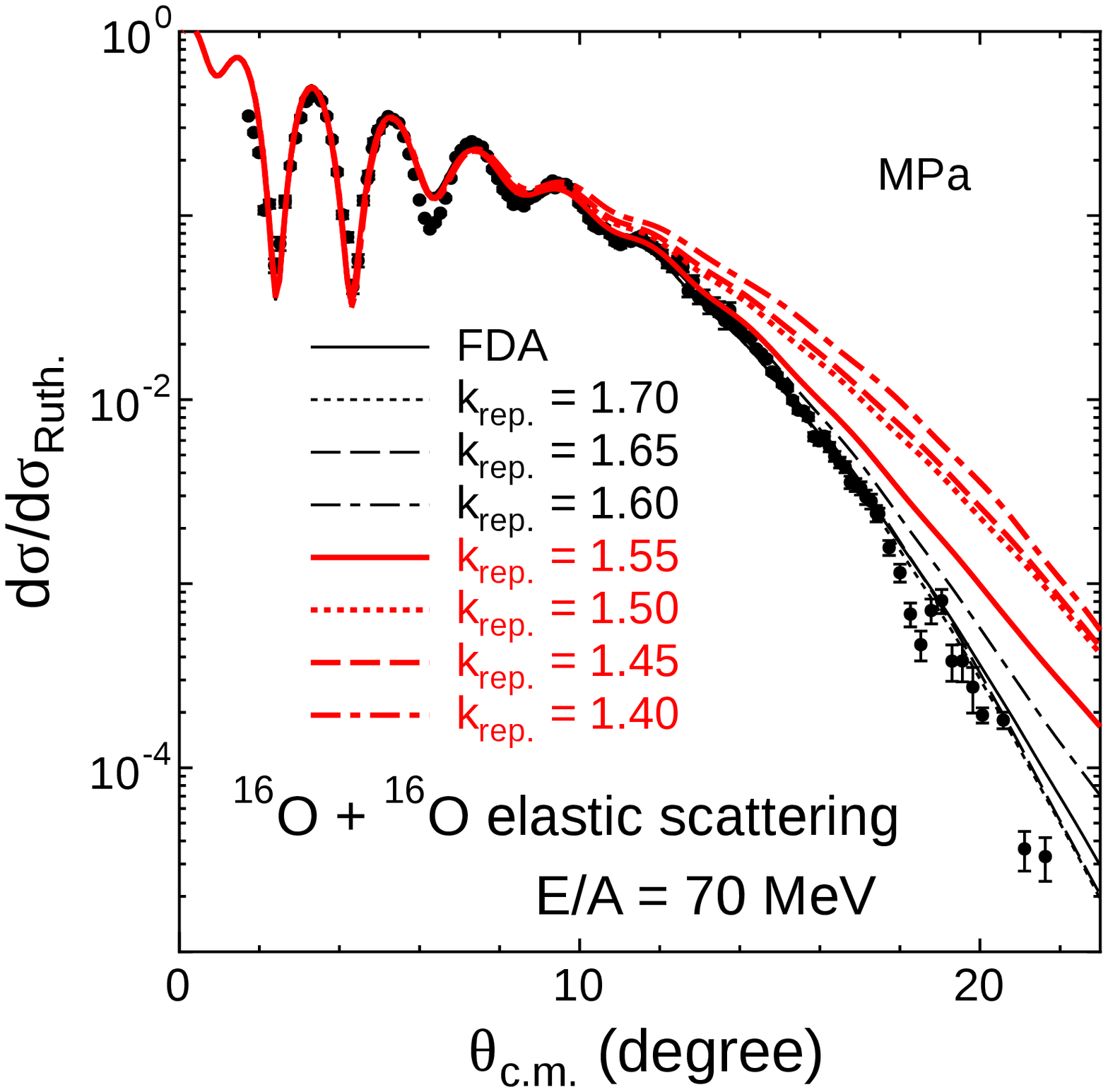}
\caption{\label{fig:04} (Color online) The elastic cross section with the DF potentials shown in Fig.~\ref{fig:03}. 
The meaning of the curves is the same as in Fig.~\ref{fig:03}.}
\end{center} 
\end{figure} 
Figure~\ref{fig:04} shows the results calculated with the DF potentials shown in Fig.~\ref{fig:03}.
The TBF effect is clearly seen up to $k_F =$ 1.65 fm$^{-1}$ in the elastic cross section while the TBF effect over $k_F =$ 1.70 fm$^{-1}$ is seen in the DF potential.
The important role of the TBF effect in the high-density region is confirmed in the nucleus-nucleus elastic cross section.
This result again implies that the nucleus-nucleus elastic scattering can sensitively probe the important role of the TBF effect in the high-density region over $k_F =$ 1.60 fm$^{-1}$.

\begin{table}[h]
\caption{The calculated total reaction cross section for the $^{16}$O~+~$^{16}$O system at $E/A =$ 70 MeV.}
\label{tab:02}
  \begin{tabular}{cc} \hline
$k_{\rm rep.}$ (fm$^{-1}$) & Total reaction cross section (mb) \\ \hline
FDA & 1428 \\ 
1.70 & 1428 \\ 
1.65 & 1428 \\
1.60 & 1428 \\ 
1.55 & 1427 \\
1.50 & 1425 \\
1.45 & 1424 \\
1.40 & 1422 \\  \hline  
  \end{tabular}
\end{table}
Next, we also calculate the total reaction cross section with the DF potentials shown in Fig.~\ref{fig:03}.
The calculated results are shown in Table~\ref{tab:02}.
The total reaction cross section has no sensitivity of the TBF effect in the high-density region.
Namely, it is difficult to discuss the detail of the TBF effect in the high-density region for the total reaction cross section.
Here, we notice that the calculated total reaction cross section is decreased by reducing the $k_{\rm rep.}$ parameter nevertheless the real and imaginary potentials become attractive.
But, this trend is not strange because the calculated total reaction cross section with the ESC interaction (w/o the TBF effect) is 1412 mb.
This value obtained by the ESC interaction is smaller than that by the MPa interaction.
Namely, the repulsive effect by the MPP and the phenomenological TBA effect in the MPa interaction give the slight increase of the total reaction cross section at this system.

\vspace{5mm}
\begin{figure}[t]
\begin{center}
\includegraphics[width=12cm]{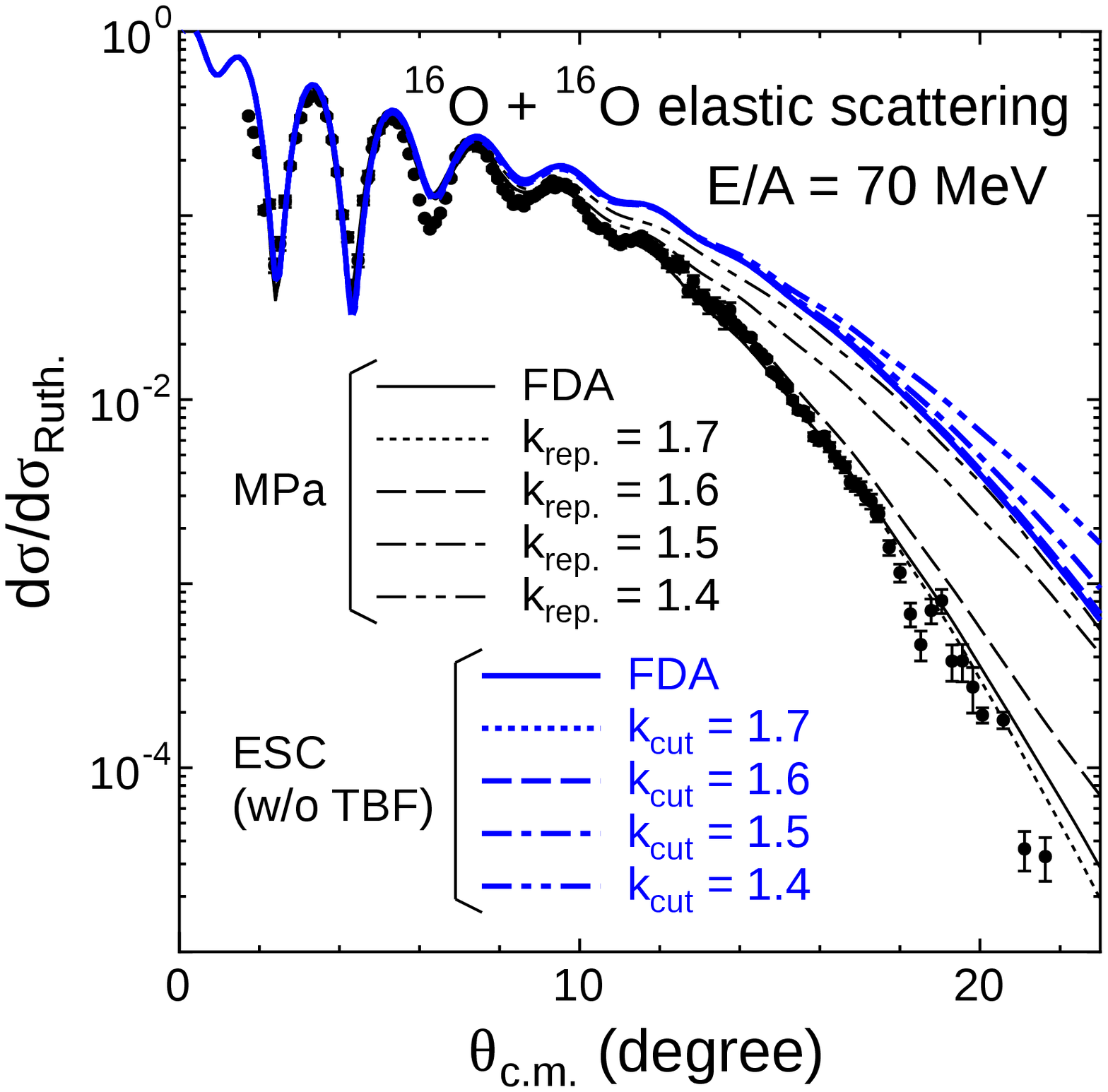}
\caption{\label{fig:05} (Color online) The elastic cross sections by the replace method based on the MPa interaction and the cutting method with the ESC interaction.}
\end{center} 
\end{figure} 
Here, we introduce the distinction of the medium effect obtained by the complex $G$-matrix calculation and the TBF effect with both of the cutting and replacement methods.
When the cutting method is applied to the ESC interaction (w/o the TBF effect), the medium effect obtained by the $G$-matrix calculation is extracted.
Next, the TBF effect included in the MPa interaction becomes clear with the replacement method.
Figure~\ref{fig:05} shows the result of distinction of the medium effects obtained by the $G$-matrix calculation and the TBF effect.
The bold (blue) curves are calculated by the cutting method with the ESC interaction (w/o the TBF effect).
The thin (black) curves are calculated by the replacement method based on the MPa interaction (with the TBF effect).
It is imperfect only by introducing the medium effect obtained by the $G$-matrix calculation to reproduce the data up to backward angles (bold (blue) solid curves).
By introducing the TBF effect, the calculated cross section around backward angles comes down and reproduces the data (thin (black) curves).
Namely, the TBF effect in the high-density region has a crucial role to reproduce the data up to the backward angles.

\subsection{Interaction dependence}
In the previous section, the MPa interaction is used to test the present methods.
We investigate the interaction dependence in this section.
We test three types of the complex $G$-matrix interactions, MPb, MPc~\cite{YAM14}, and CEG07b~\cite{FUR09} and one type of the effective density-dependent $NN$ interaction, CDM3Y6~\cite{KHO97}.
The MPa interaction includes the triple- and quadruple-pomeron exchange contributions.
The quadruple-pomeron exchange contribution is not included in the MPb and MPc interactions.
Namely, the TBF effect of the MPb and MPc interactions is described only by the three-body pomeron exchange contribution while their strength is different.
The different MPP contribution has a minor role in the heavy-ion elastic scattering although the contribution is clearly seen in EOS of the neutron star \cite{YAM14,YAM16}.
As mentioned in Ref.~\cite{FUR14R}, other complex $G$-matrix interactions are out of range for the FDA calculation because they does not have the density dependence up to twice the normal density.
Here, the values of incompressibility $K$ for the present $G$-matrix interactions, MPa, MPb, MPc and CEG07b are obtained as 283 MeV at $\rho_0=0.154$ fm$^{-3}$, 254 MeV at $\rho_0=0.154$ fm$^{-3}$, 243 MeV at $\rho_0=0.156$ fm$^{-3}$, and 197 MeV at $\rho_0=0.160$ fm$^{-3}$, respectively.
The CEG07b interaction gives the small value of incompressibility because the saturation energy (the $E/A$ value at the saturation density) in this case is obtained as $-14.2$ MeV~\cite{WWQ15} considerably shallower than the values of $\sim -16$ MeV for MPa, MPb, and MPc.
The DF potential for CEG07b reproduce well the several nucleus-nucleus elastic scatterings~\cite{FUR09} as well as MPa, MPb and MPc, though their derived saturation properties
are different from each other. 
The saturation curves are obtained by summing up $G$-matrices for all nucleon pairs in nuclear matter, while DF potentials are by $G$-matrices between a positive-energy nucleon and other nucleons.
There is no direct connection between the saturation curve and the DF potential, being calculated independently.
Then, there still remain ambiguities in the TBF part for saturation properties, even if they are chosen so as to give reasonable DF potentials.
This imply that the correct reproduction of the saturation property becomes the precondition for finding some linkage between the heavy-ion elastic scattering and the incompressibility.
In this concern, the incompressibility obtained with the CDM3Y6 interaction is obtained as 252 MeV at $\rho_0=0.17$ fm$^{-3}$~\cite{KHO07}.
For comparison, we evaluate the $K$ values at 0.17 fm$^{-3}$ for MPa, MPb and MPc, being 343 MeV, 305 MeV and 283 MeV, respectively.
Namely, the incompressibilities for these interactions turn out to be substantially larger than that for CDM3Y6, which corresponds to the difference between the $^{16}$O~+~$^{16}$O elastic cross sections at backward angles.
It should be noted that the $K$ value for MPa is larger than that for MPb (MPc) due to the quadruple-pomeron exchange contribution included in MPa, and the difference between MPa and MPb (MPc) is quite small in the the $^{16}$O~+~$^{16}$O elastic cross sections.

The symmetry energy $E_{sym}$ and the slope parameter $L$ for the MPa (MPb) interaction are obtained as 31.7 (31.7) and 67.1 (66.7) MeV~\cite{HIY16}.
These values are not so different from those obtained from ESC08c, because the TBF parts in MPa (MPb) are not dependent on iso-spins.
There are many trials to extract the experimental constraints for $E_{sym}$ and $L$.
As an example, these values can be compared to those in Ref.~\cite{LAT14}:
Our values of $E_{sym}$ are of nice correspondence to the experimental indication, and our $L$ values are near to the upper limit of the experimental one.
Anyway, the values of $E_{sym}$ and $L$ are related not directly to the DF potentials.

Here, we introduce the renormalization factor in the case with the complex $G$-matrix interaction as follows;
\begin{equation}
U = V + N_W W,
\end{equation}
where $V$ and $W$ mean the real and imaginary parts of the DF potential, respectively.
Namely, the renormalization factor is a parameter to change the imaginary strength.
On the other hand, the CDM3Y6 interaction has no imaginary part.
Then, we assume the imaginary part as follows;
\begin{equation}
U = (1 + i N_W) V, \label{eq:recdm3y}
\end{equation}
where, $V$ is the real part of the DF potential with the CDM3Y6 interaction.
Namely, the real and imaginary parts of the potential are assumed to have the same form in the case with the CDM3Y6 interaction.

\begin{figure}[t]
\begin{center}
\includegraphics[width=12cm]{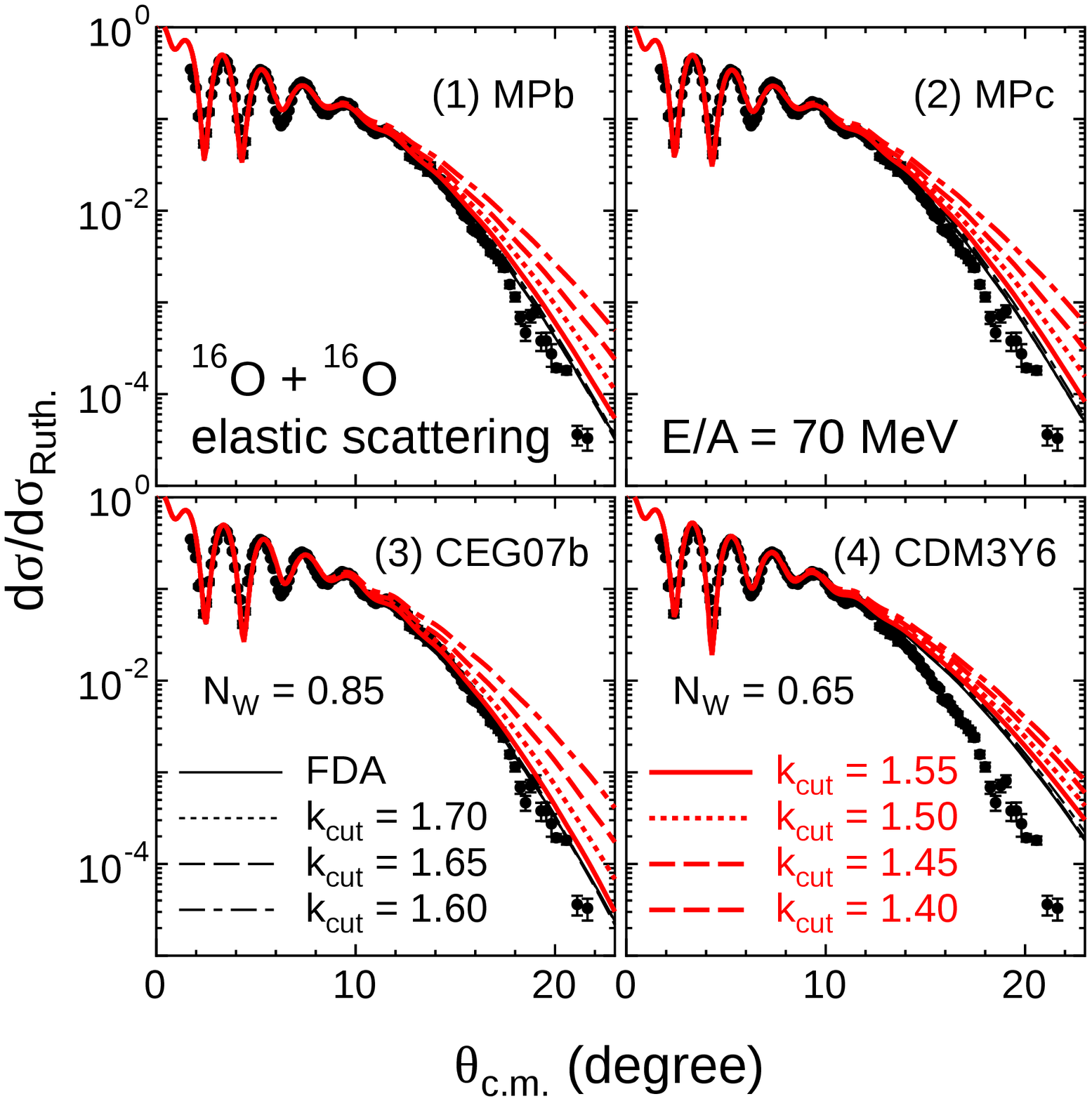}
\caption{\label{fig:06} (Color online) Same as Fig.~\ref{fig:02} but with the MPb, MPc, CEG07b, and CDM3Y6 interactions.}
\end{center} 
\end{figure}
Figure~\ref{fig:06} shows the results of the cutting method with the MPb, MPc, CEG07b, and CDM3Y6 interactions.
The results with the MPb and MPc interactions need no renormalization factor in the same case with the MPa interaction.
The renormalization factors for the DF potential with the CEG07b and CDM3Y6 interactions are fixed to be 0.85 and 0.65, respectively.
The medium effect in the high-density region is clearly seen in the elastic cross section up to $k_F = 1.60$ fm$^{-1}$.
The repulsive TBF effect of the MPa, MPb, and MPc interactions is obtained by the MPP, and their attractive TBF effect is introduced by the phenomenological way~\cite{YAM14}.
On the other hand, the TBF effect of the CEG07b interaction is obtained by changing the vector meson mass, and the attractive TBF effect is described by the Fujita-Miyazawa diagram~\cite{FUR09}.
Namely, the modeling of the TBF is completely different.
Nevertheless, the behaviors of the medium effect in the high-density region are similar to each other while the renormalization factor is introduced.
In addition, the medium effect by the phenomenological density-dependent effective $NN$ interaction (CDM3Y6) also gives similar result but the data is not reproduced up to backward angles when we assume the strength and form of the complex potential as Eq.(\ref{eq:recdm3y}).

\begin{figure}[t]
\begin{center}
\includegraphics[width=12cm]{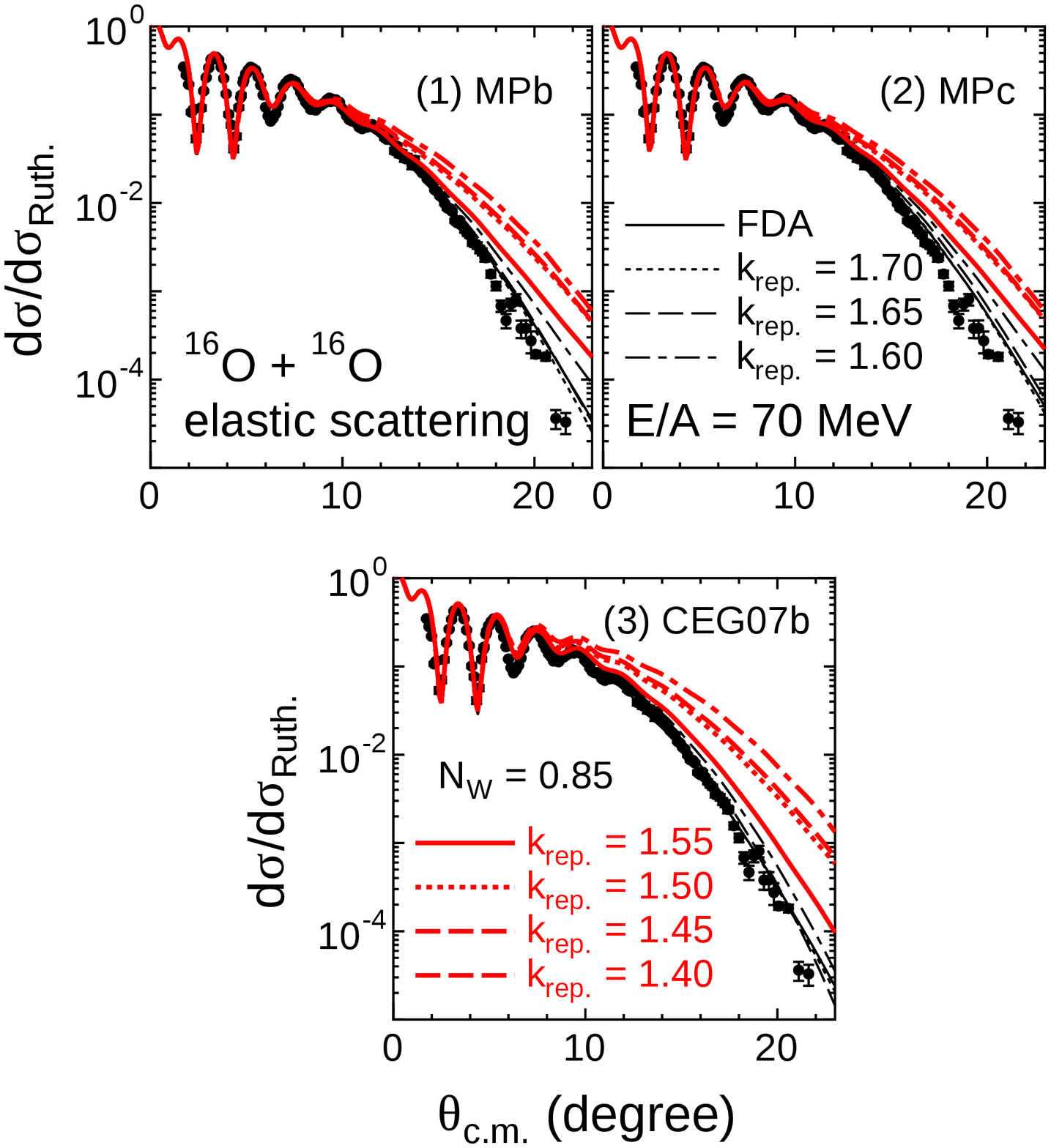}
\caption{\label{fig:07} (Color online) Same as Fig.~\ref{fig:04} but based on the MPb, MPc, and CEG07b interactions.}
\end{center} 
\end{figure}
Next, we test the replacement method based on the MPb, MPc, and CEG07b interactions.
The results are shown in Fig.~\ref{fig:07}.
The TBF effect is clearly seen on the elastic cross section over $k_F = 1.6$ fm$^{-1}$.
Here, we use the CEG07a interaction (w/o the TBF effect) instead of the ESC interaction in the case with CEG07b because the CEG07a and CEG07b interactions are based on the same $NN$ interaction, ESC04.
While the modeling of the TBF effect for the MPa and CEG07b interaction is different, the effect emerged by the replacement method gives the similar behaviors.
Because the effect of the interaction dependence is not so large with the present methods as shown in this section, we apply only the MPa and ESC interactions to other analyses.

\subsection{Target mass dependence}
The $^{16}$O~+~$^{16}$O system is one of ideal systems to investigate several information from the elastic scattering because the $^{16}$O nucleus is one of the magic nucleus and has no collective excited states strongly coupled to the ground state.
However, we apply the present methods to the several systems in this section.
The $^{16}$O nucleus is located in the small and light region for the table of nuclides.
We here focus on the effects of the nuclear radius and form by changing the target nucleus.
From this section, we present the combined analysis as shown in Fig.~\ref{fig:05}.

\begin{figure}[t]
\begin{center}
\includegraphics[width=12cm]{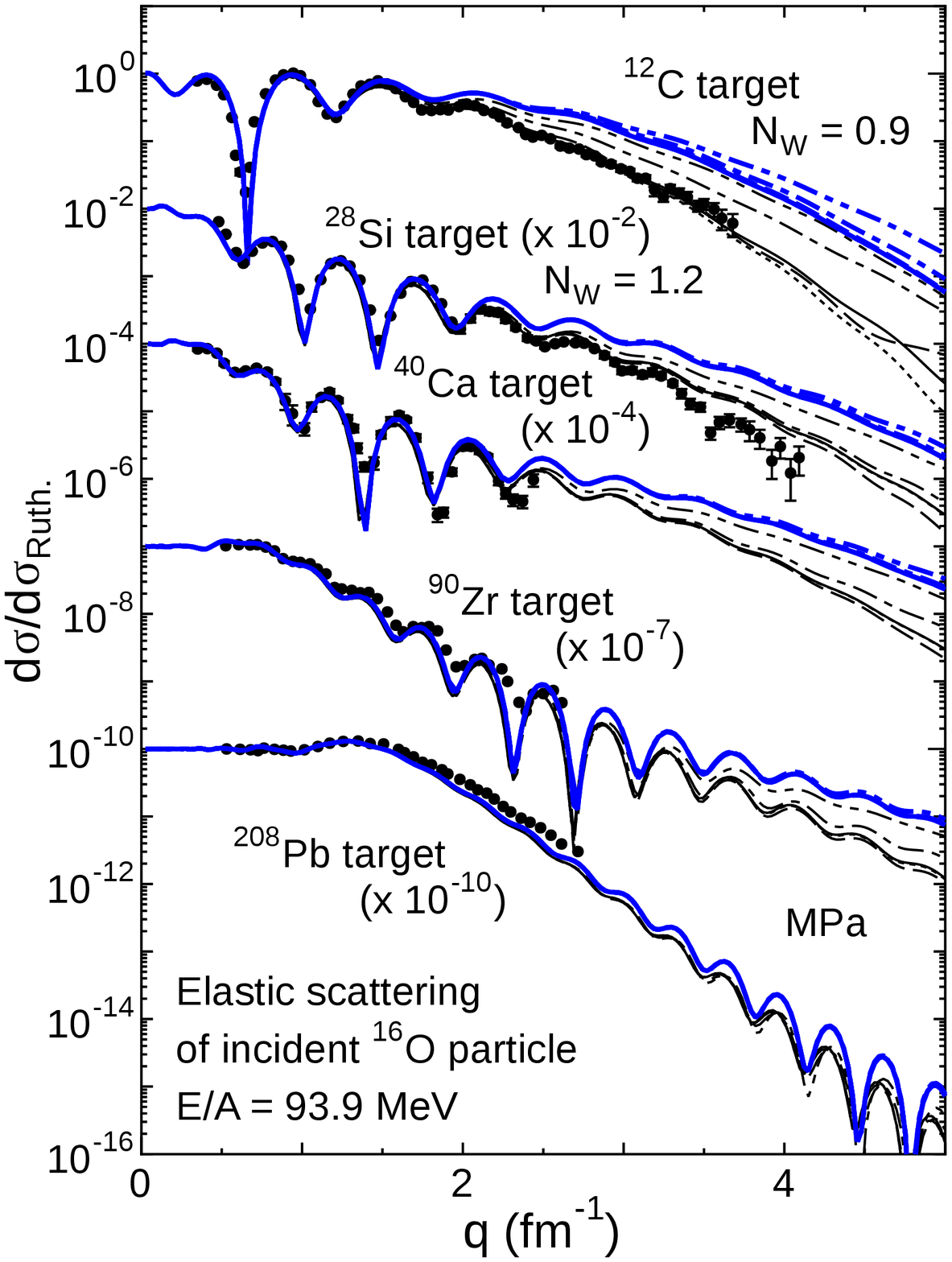}
\caption{\label{fig:08} (Color online) Elastic cross section for the incident $^{16}$O particle by the several targets at $E/A =$ 93.9 MeV.
The meaning of the curves is the same as in Fig.~\ref{fig:05}.
It means $N_W = 1$ without the $N_W$ value.
The experimental data is taken from Ref.~\cite{ROU88}.
}
\end{center} 
\end{figure}
We test the present replacement method based on the MPa interaction and the cutting method with the ESC interaction for the incident $^{16}$O nucleus at $E/A =$ 93.9 MeV to investigate the target mass dependence.
The results are shown in Fig.~\ref{fig:08}.
The medium effects obtained by the $G$-matrix calculation and the TBF effect are clearly seen in the backward angles but the target mass and incident energy are changed.
When the target mass becomes larger and larger, it seems that the medium effects obtained by the $G$-matrix calculation and the TBF effect in the high-density region becomes small.
Because it is considered that the imaginary part of the DF potential also becomes large with increasing the target mass.
In addition, the Coulomb potential strength increases by changing the target nucleus.
It is difficult to investigate the inner part of the potential in detail by the increase of the imaginary and Coulomb potentials.
This result implies that the present method prefers the small system to large one.

\subsection{Incident energy dependence}
\begin{figure}[t]
\begin{center}
\includegraphics[width=12cm]{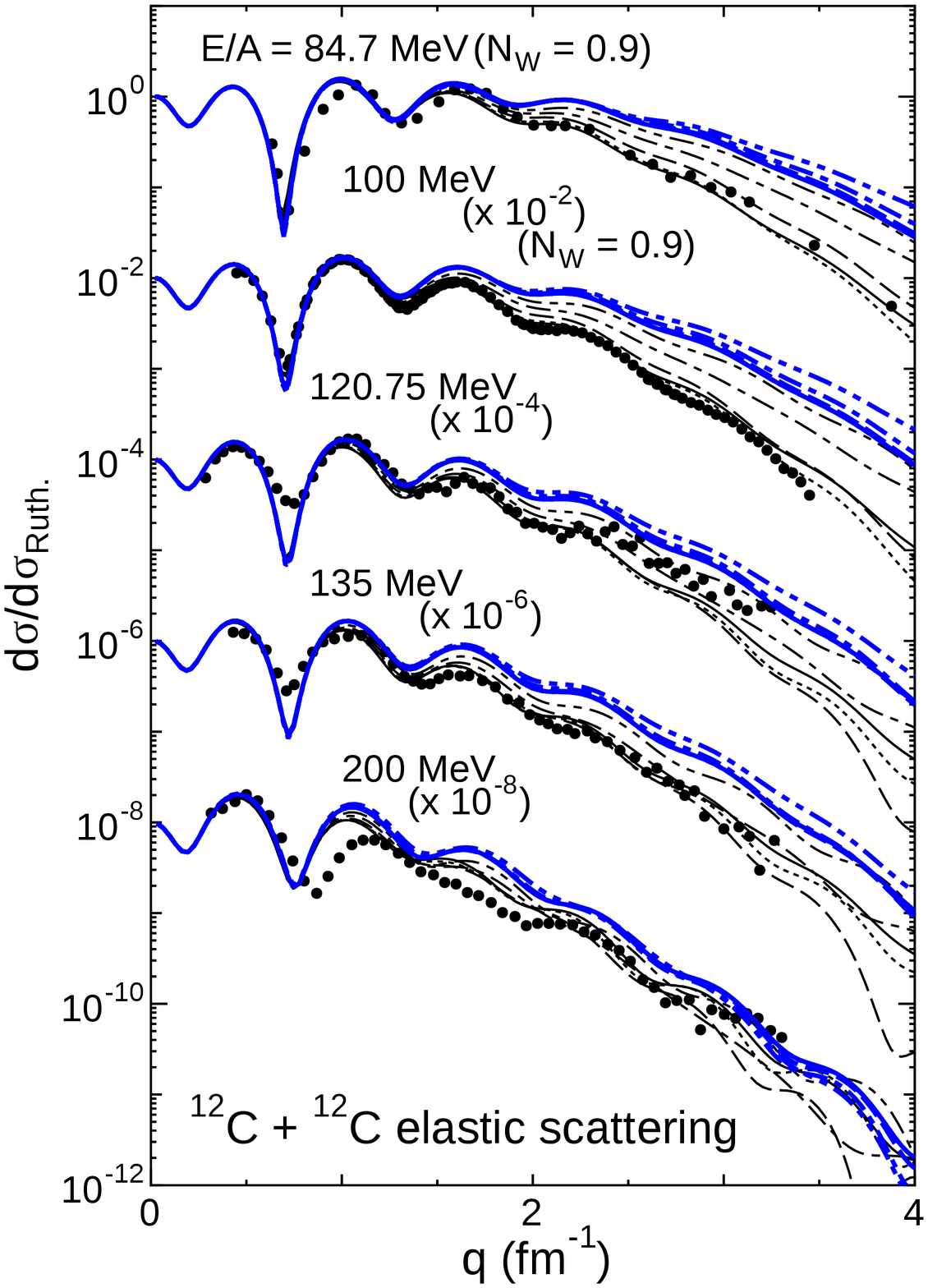}
\caption{\label{fig:09} (Color online) Elastic cross section for the $^{12}$C~+~$^{12}$C system at the various incident energies.
The meaning of the curves is the same as in Fig.~\ref{fig:05}.
The experimental data is taken from Refs.~\cite{BUE81,WWQ15,ICH94,HOS87}.
}
\end{center} 
\end{figure}
Next, we focus on the energy dependence.
Figure~\ref{fig:09} shows the results of the replacement method based on the MPa interaction and the cutting method with the ESC interaction for the $^{12}$C~+~$^{12}$C system at $E/A =$ 84.7--200 MeV.
When the incident energy becomes larger and larger, it is difficult to distinguish the medium effect in the high-density region.
The effect by the replacement method at the high-incident energy slightly remains around the most backward angles.
However, the experimental data does not exist around the most backward angles.
There are two reasons for this minor medium effect in the high-density region around $E/A =$ 200 MeV.
First, the imaginary part of the DF potential in the high-energy region is larger than that in the low-energy region.
Second, it is predicted that the real part of the folding potential around $E/A =$ 200--300 MeV close to be 0, while with the CEG07b interaction~\cite{FUR10}.
This implies that the real part of the potential has small effect on the elastic scattering in this energy range.
Therefore, the medium effect in the high-density region is not clearly seen in the elastic cross section at $E/A =$ 200 MeV. 
Here, we should note that the medium effect is clearly seen in the elastic cross section when we set the $k_{\rm cut}$ or $k_{\rm rep.}$ values to be below 1.4 fm$^{-1}$.
Because the purpose in this paper is to investigate the medium effect over the normal density, the $k_{\rm cut}$ or $k_{\rm rep.}$ values are selected over 1.4 fm$^{-1}$.
In addition, we noted that the real potential is close to 0 when the incident energy per nucleon increases. 
However, it is expected that the real part of the DF potential again has an influence to the elastic scattering reaction if we select the incident energy higher than the 200 MeV.
Then, we test the medium effect in the high-density region for the $^{12}$C + $^{12}$C system up to $E/A =$ 400 MeV.

\begin{figure}[t]
\begin{center}
\includegraphics[width=12cm]{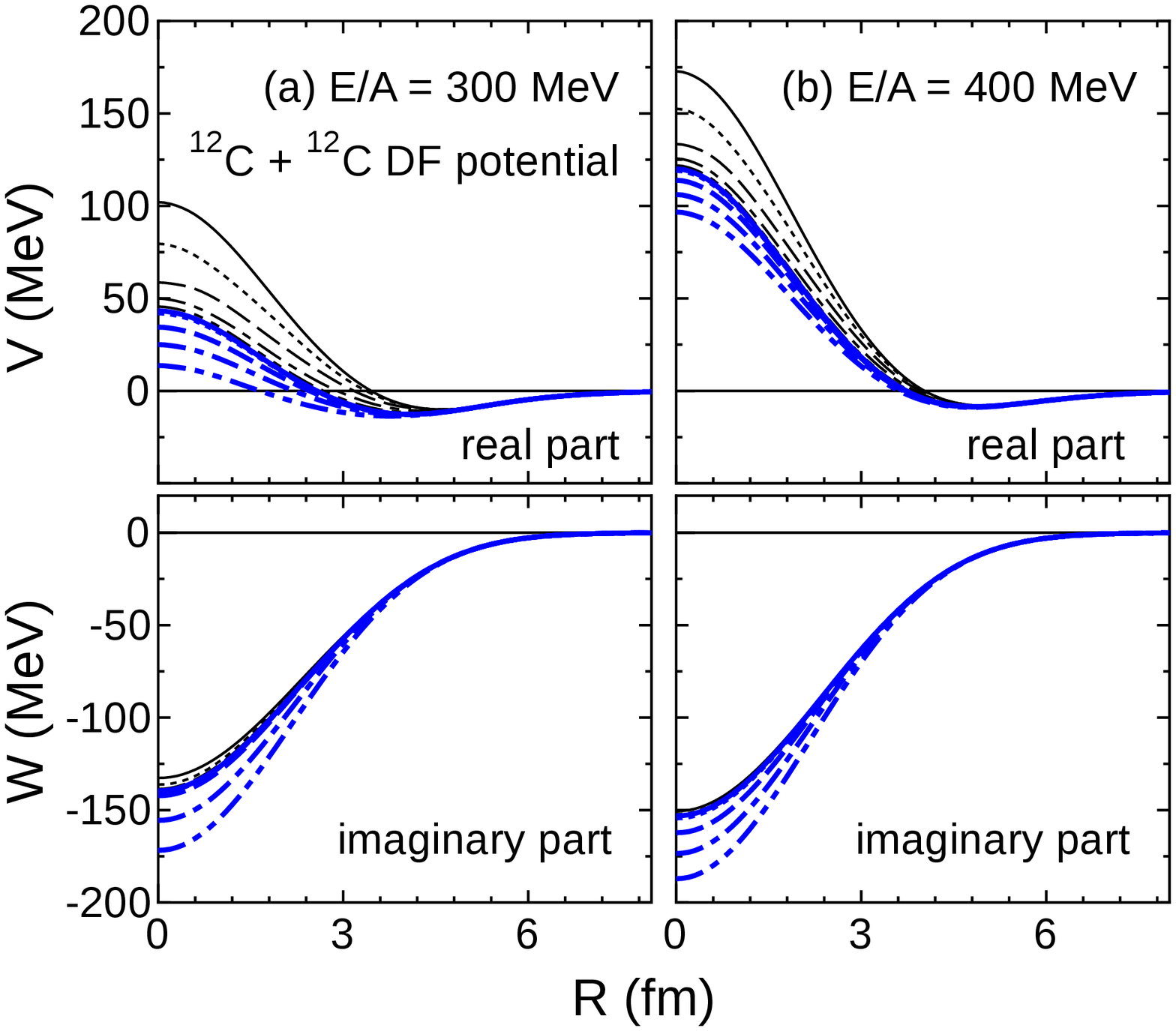}
\caption{\label{fig:10} (Color online) The real (upper panel) and imaginary (lower panel) parts of the DF potential at $E/A =$ 300 and 400 MeV. 
The meaning of the curves is the same as in Fig.~\ref{fig:05}.}
\end{center} 
\end{figure}
Figure~\ref{fig:10} shows the DF potentials obtained by the replacement method based on the MPa interaction and the cutting method with the ESC interaction.
The medium effect obtained by the $G$-matrix calculation with the cutting method is clearly seen in the both of the real and imaginary parts.
The TBF effect is clearly seen in the real part of the DF potential.
On the other hand, the TBF effect on the imaginary part of the DF potential is invisible.
With switching off the medium effect in the high-density region by the replacement and cutting methods, the real part of the DF potential show the change from repulsion to attraction.

\begin{figure}[t]
\begin{center}
\includegraphics[width=12cm]{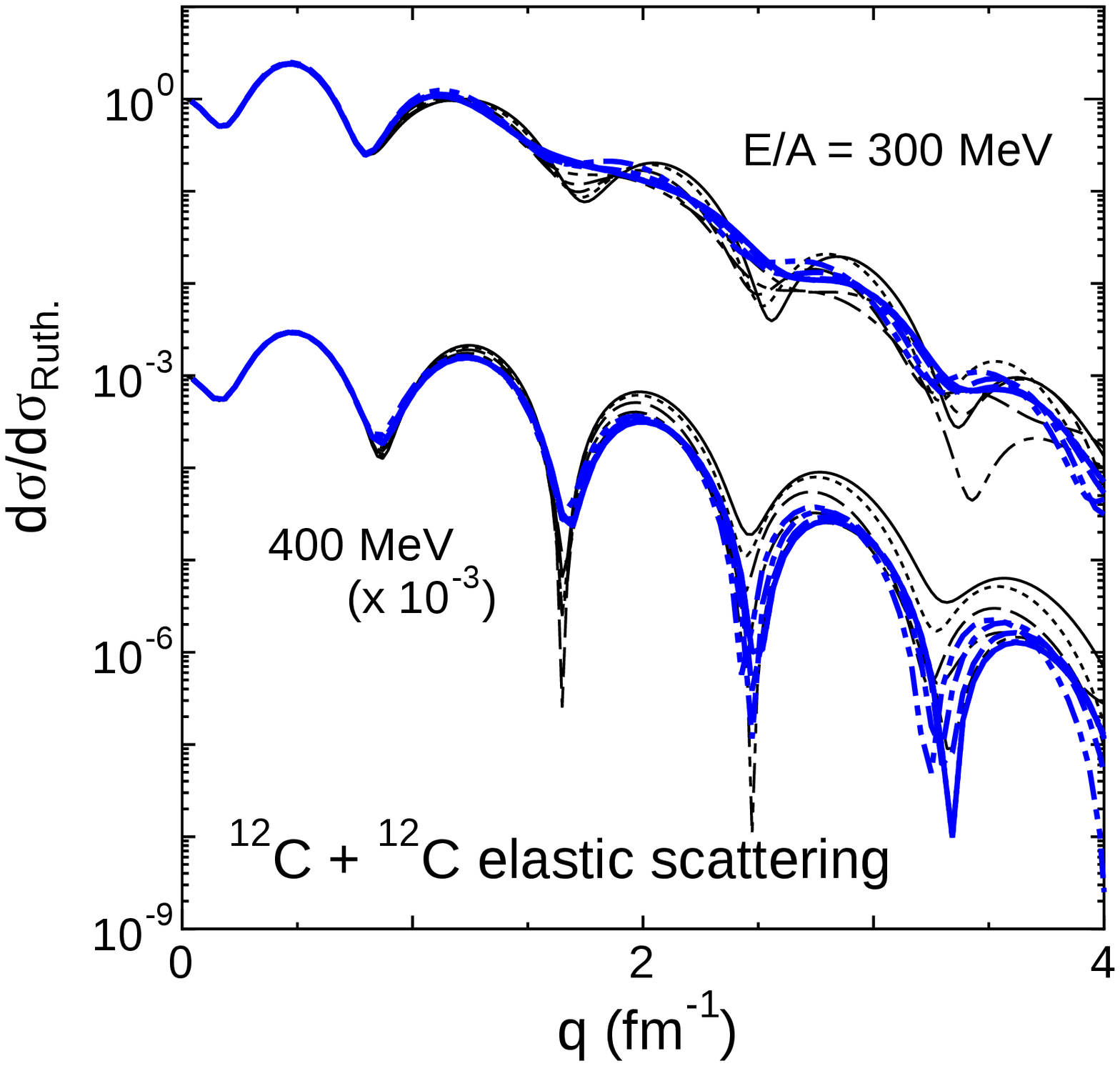}
\caption{\label{fig:11} (Color online) The elastic cross section with the DF potentials shown in Fig.~\ref{fig:10}. 
The meaning of the curves is the same as in Fig.~\ref{fig:05}.}
\end{center} 
\end{figure} 
We apply the DF potentials shown in Fig.~\ref{fig:10} to the high-energy heavy-ion elastic scattering.
Figure~\ref{fig:11} shows the result for the $^{12}$C~+~$^{12}$C system at $E/A =$ 300 and 400 MeV.
By the replacement method, the TBF effect in the high-density region is clearly seen on the high-energy scatterings.
At $E/A = 300$ MeV, the effect of the shift of the k$_{\rm rep.}$ value is not smooth because the folded real potential around $R = 3$ fm shows the transition from repulsion to attraction.
The effect of the cutting method described by the bold (blue) curves is not shown on the elastic cross section.
Because the real part of the potential at 300 MeV is too weak to affect on the elastic cross section as well as the imaginary strength is large enough even without the cutting method.
At $E/A = 400$ MeV, the folded real potential is sufficiently repulsive.
Then, the effect of the inner part of the real potential well appears in the elastic cross section.
By the cutting method, the calculated elastic cross sections at 400 MeV show the change as the reverse of the effect by the replacement method.
This change is caused by the transition from repulsion to attraction as shown in Fig.~\ref{fig:10}.
Finally, the cutting and replace methods are the reliable tool to investigate the medium effect over the normal density even for the high-energy heavy-ion elastic scattering.
However, we need the careful treatment because the transition from repulsion to attraction of the real part of the optical potential is shown in the high-energy elastic scattering.

\section{Summary and Remarks}
In summary, we have constructed the DF potential with the complex $G$-matrix interaction including the TBF effect based on the MPP model.
The medium effect in the high-density region have been investigated with two methods which are called as ``cutting method'' and ``replacement method''.
With both methods, the $^{16}$O~+~$^{16}$O elastic scattering at $E/A =$ 70 MeV is investigated.
The medium effect including the TBF effect in the region over the normal density is clearly seen in the potential and the elastic scattering cross section.
In addition, we made clear that the TBF effect up to $k_F =$ 1.6 fm$^{-1}$ has a critical role to determine the heavy-ion elastic angular distribution.

In order to confirm our result, the interaction dependence is also investigated.
The similar results are obtained with other complex $G$-matrix interactions, MPb/c and CEG07b.
In this paper, we test not only the complex $G$-matrix interaction but also the phenomenological density-dependent interaction, CDM3Y6.
The CDM3Y6 interaction also gives the similar result, and it shows the important role of the medium effect in the high-density region.
Again, we made clear that the TBF effect up to $k_F =$ 1.6 fm$^{-1}$ has a critical role to determine the elastic angular distribution.

In addition, the target mass and the incident energy dependences are investigated.
The target mass dependence is investigated for the incident $^{16}$O particle by the $^{12}$C, $^{28}$Si, $^{40}$Ca, $^{90}$Zr, and $^{208}$Pb  targets at $E/A =$ 93.9 MeV.
When the target mass number becomes larger and larger, the medium effect obtained by both of the $G$-matrix calculation and the TBF effect becomes invisible.
Because the imaginary and Coulomb potentials also become large.
On the other hand, the medium effect in the high-density region is clearly seen for the $^{12}$C~+~$^{12}$C elastic scattering at various incident energies.
However, it becomes difficult to see the medium effect at $E/A = 200$ MeV on the elastic cross section.
In this case, not only the increase of the imaginary strength but also the decrease of the real strength is contributed.
However, the cutting and replacement methods again becomes powerful tool to investigate the medium effect over the normal density when we increase the incident energy over 200 MeV.

Our results imply that the medium effect including the TBF effect in the high-density region can be probed by the observed nucleus-nucleus elastic scattering.
Finally, we made clear the crucial role of the TBF effect in the high-density region on the nucleus-nucleus elastic scattering.

\vspace{2mm}
\section{Acknowledgment}
The authors acknowledge Professor Uesaka for encouraging comments.
This work was supported by JSPS KAKENHI Grant Numbers 15H00842 and 15K05087.


\end{document}